\documentclass[]{pasj01}

\begin{document} 
\Received{}
\Accepted{}

\title{X-ray upper limits of GW151226 with MAXI}

\author{Motoko~\textsc{Serino}\altaffilmark{1}}%
\altaffiltext{1}
{MAXI team, RIKEN, 2-1 Hirosawa, Wako, Saitama 351-0198, Japan}
\email{motoko@crab.riken.jp}

\author{Nobuyuki~\textsc{Kawai}\altaffilmark{2,1}}
\altaffiltext{2}
{Department of Physics, Tokyo Institute of Technology, 
2-12-1 Ookayama, Meguro-ku, Tokyo 152-8551, Japan}

\author{Hitoshi~\textsc{Negoro}\altaffilmark{3}}
\altaffiltext{3}{Department of Physics, Nihon University, 
1-8-14 Kanda-Surugadai, Chiyoda-ku, Tokyo 101-8308, Japan}


\author{Tatehiro~\textsc{Mihara}\altaffilmark{1}}%

\author{Takahiro~\textsc{Masumitsu}\altaffilmark{3}}%

\author{Satoshi~\textsc{Nakahira}\altaffilmark{4}}%
\altaffiltext{4}
{JEM Mission Operations and Integration Center, 
Human Spaceflight Technology Directorate, 
Japan Aerospace Exploration Agency, 
2-1-1 Sengen, Tsukuba, Ibaraki 305-8505, Japan}


\maketitle

\begin{abstract}
The error region of the the gravitational-wave (GW) event GW151226 
was observed with Monitor of All-sky X-ray Image (MAXI).
MAXI was operated at the time of GW151226, and continuously observed
to 4 minutes after the event.
MAXI covered about 84\% of the 90 percent error region of 
the GW event during the first 92 minutes orbit after the event.
No significant X-ray transient was detected in the GW error region.
A typical 3-$\sigma$ GSC upper limit for a scan is 
1.2 $\times 10^{-9}$ ergs cm$^{-2}$ s$^{-1}$ in the 2--20 keV.
The auto-detection (MAXI nova-search) systems detected a short excess 
event with a low significance (2.85$\sigma$)
from 5257 s to 5260 s after the GW trigger.
Finally, we discuss the sensitivity of MAXI to long X-ray emissions 
of short gamma-ray bursts, 
which are expected to accompany GW events.
\end{abstract}

\section{Introduction}

The second detection of the gravitational wave (GW) event was made
by LIGO (Laser Interferometer Gravitational-Wave Observatory)
on 2015 December 26, 03:38:53.647 UT, and named GW151226
\citep{2016PhRvL.116x1103A}.
The event was a merger masses of two black holes (BHs),
of which the masses are estimated 
as 14.2$^{+8.3}_{-3.7}$ and 7.5$^{+2.3}_{-2.3}$ $M_{\odot}$. 
A luminosity distance of 440$^{+180}_{-190}$ Mpc was derived by the 
waveform analysis.
An important aspect of this event is
that it added another example of a merger of BHs
with the total mass of $> 20M_{\odot}$ after the first detection
of GW150914
\citep{2016PhRvL.116f1102A}.

These GW detection motivated theoretical studies of electromagnetic (EM)
radiations from BH mergers
\citep{2016PTEP.2016e1E01Y,2016ApJ...819L..21L,2016ApJ...821L..18P,
2017MNRAS.465.4406K}.
Merger events are possible sources of short gamma-ray bursts
(SGRBs;
\cite{2007PhR...442..166N,2014ApJ...796...13N}),
which emit gamma-rays and X-rays for a few -- several hundred seconds.
There are models which predict that
these emissions arise from relativistic jets or outflows.
EM counterparts can provide essential information of sources
such as host galaxies and populations of nearby stars.
Searches of an X-ray counterpart of this GW event
were carried out by the Fermi/GBM \citep{2016arXiv160604901R},
the CALET Gamma-ray Burst Monitor \citep{2016ApJ...829L..20A},
Swift \citep{2016MNRAS.462.1591E}, XMM-Newton
\citep{2016arXiv160604901R}, and 
Monitor of All-sky X-ray Image (MAXI; \cite{2009PASJ...61..999M}).
No X-ray counterpart was reported for this GW event by any of them,
which was similar to the result of GW150914
\citep{2016ApJ...826L..13A}.

MAXI is an instrument for X-ray monitor
from the International Space Station (ISS). 
It scans most of the sky every orbits ($\sim$ 92min) of the
ISS with a narrow and long field of view. 
MAXI searched for an X-ray counterpart of GW150914,
and reported upper limits
\citep{2016GCN..19013...1S,gw150914.maxi}.
In this paper we present detailed results of the MAXI follow-up 
of GW151226,
following a quick report of MAXI non-detection by
\citet{2015GCN..18784...1N}.

\section{Instrumentation}
MAXI on ISS
has two instruments: the GSC (Gas Slit Camera; \cite{2011PASJ...63S.623M})
and the SSC 
(Solid-state Slit Camera; \cite{2011PASJ...63..397T}).
The field of views (FOVs) of GSC is about 2 \% of the whole sky.
The GSC is not operating in the regions with high particle-background,
including the South Atlantic Anomaly and regions with the
latitude higher than $\sim 40$ 
degrees, and in the vicinity of the sun ($\sim 5$ degrees).
Although the operating duty ratio is about 40\%,
the GSC covers about 85 \% of the whole sky in a scan
\citep{2011PASJ...63S.635S}.
Because the SSC is operated in the night time to avoid the sun light,
its operating efficiency becomes considerably low.
The SSC duty ratio and sky coverage is about 25-30\% and 30\% 
respectively.

A nominal GSC camera can detect transient events with 
the 2--20 keV flux 
of $\sim$2 $\times 10^{-9}$ erg cm$^{-2}$ s$^{-1}$ 
(e.g. \cite{2014PASJ...66...87S,2016PASJ...68S...1N}) in a scan transit.
Seven out of the 12 GSC cameras were functioning 
at the time of GW151226.
However, one GSC camera (GSC\_0) has a gas leak and lost
the stopping power in the high energy band 
\citep{2014SPIE.9144E..1OM}.
In addition, 
two GSC cameras (GSC\_3 and GSC\_6) are working with a half effective
area due to the loss of a carbon anode wire, and suffer high background 
because of loss of the veto cells at the bottom of the counters.
Since GSC\_0 and GSC\_6 observe the same sky region and GSC\_0 has
better sensitivity
(even though the high energy efficiency is less), 
we did not use the data of GSC\_6.

\section{Observations}
 \subsection{the observation and coverage of GSC and SSC}

\begin{figure}
 \begin{center}
 \includegraphics[width=8cm]{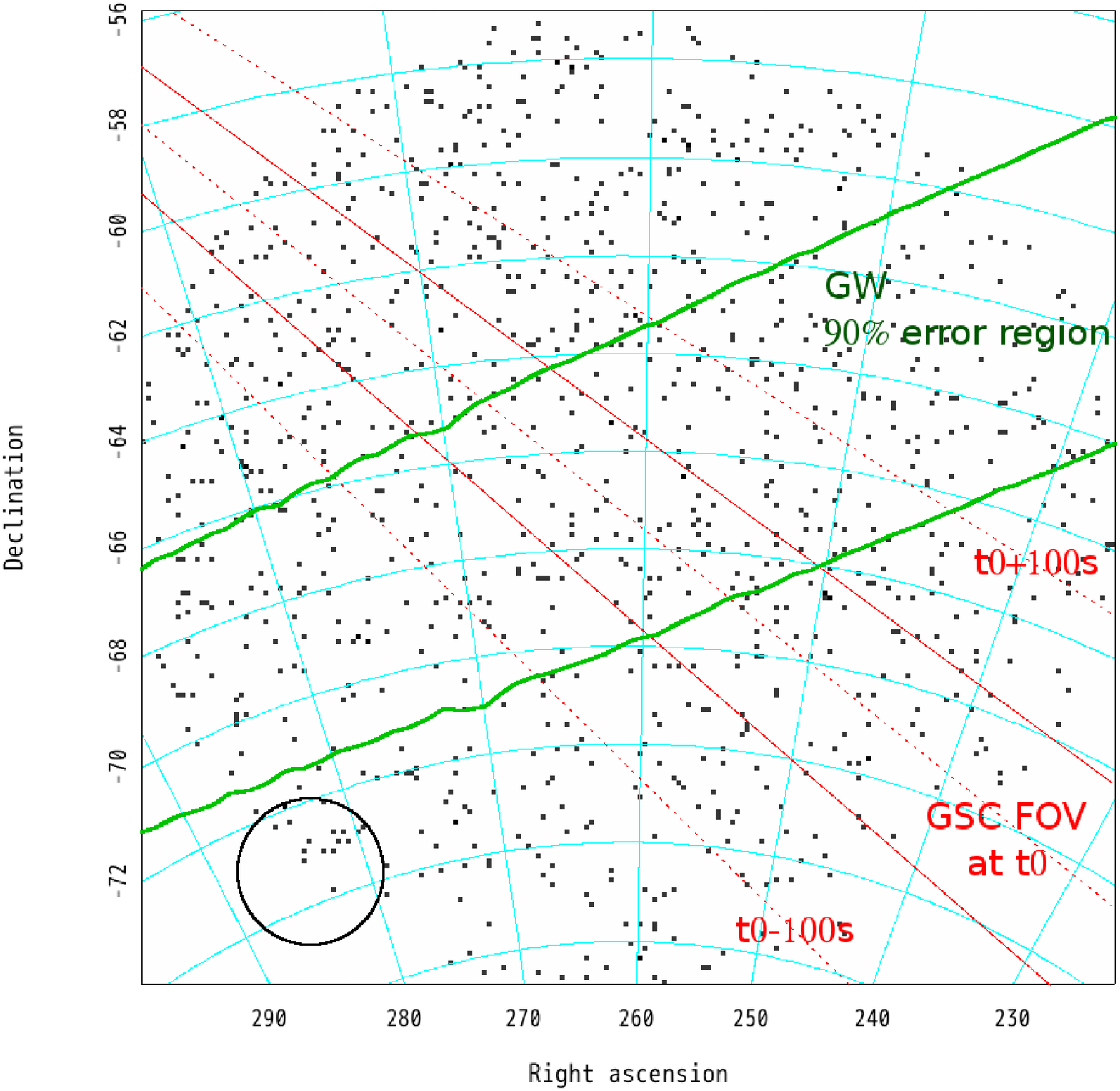}
 \end{center}
\caption{An X-ray event map around the GW error region observed by GSC from 
  $t0-$269 s to $t0$+263 s.
The BAYESTAR 90\% probability region is between the (green) solid curves.
 GSC scanned from left bottom to right top in this image.
 The dashed lines show the center of the GSC FOV at $t0-100$ s, $t0$,
 and $t0+100$ s. The solid lines are the edges of the GSC FOV at $t0$.
 The black points are X-ray photons.
 A circle at the left bottom shows 
 a typical size of the point spread function of GSC (3 deg diameter).
 The image is in the equatorial coordinates.
 (Color online)
      }
\label{fig:gsc100s}
\end{figure}

\begin{figure*}
 \begin{center}
 \includegraphics[width=12cm]{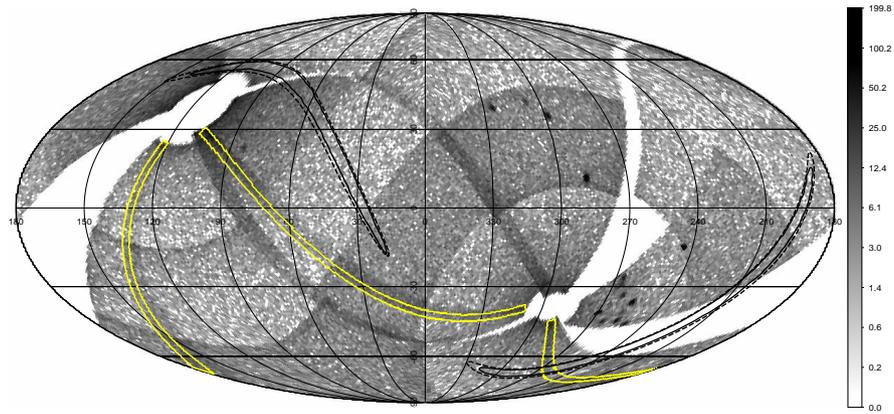}
 \end{center}
\caption{An X-ray image observed by GSC from 
  $t0-$100 s to $t0$+92 minutes.
The GW 90\% probability contours of
 the LALInference and the BAYESTAR are shown in solid and dashed lines,
 respectively.
 The white (yellow in the color version)
 regions are the FOV of GSC at the GW event.
 The unit of gray-scale is counts per pixel of HEALPix.
 The image is in the equatorial coordinates.
 (Color online)
      }
\label{fig:gsc1scan}
\end{figure*}

  The GSC was operated at the time of GW event 
  ($t0$ = 2015-12-26T03:38:53.648 UT). 
  The FOVs of the GSC at the time
  is indicated as white (yellow in the color version) lines
  in figure \ref{fig:gsc1scan}.
  Because the instantaneous FOVs of GSC is only 2\% of the whole sky,
  it covered, at the time of the event,
  1\% of the 90 percent error region of the GW location.
  Figure \ref{fig:gsc100s} shows only the part that GSC covered
  the GW error region.
  There is no count excess along the line of $t0$ or in the
  GW error region, 
  which means no obvious X-ray counterpart of short timescale
  around the time of the GW event.

\begin{figure*}
 \begin{center}
 \includegraphics[width=12cm]{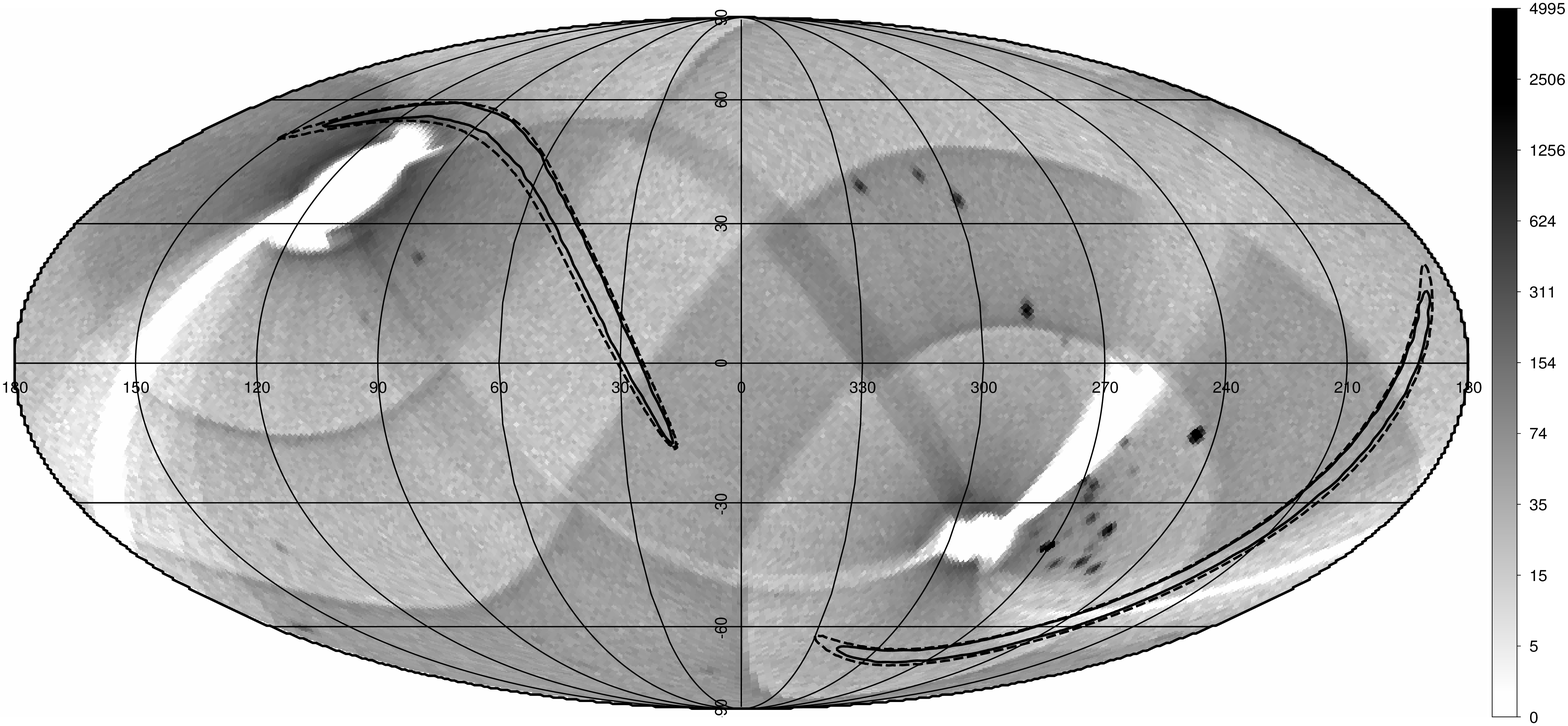}
 \end{center}
\caption{Same as figure \ref{fig:gsc1scan} but 
  from $t0$ to $t0$+1day.
      }
\label{fig:gsc1day}
\end{figure*}

\begin{figure*}
 \begin{center}
 \includegraphics[width=12cm]{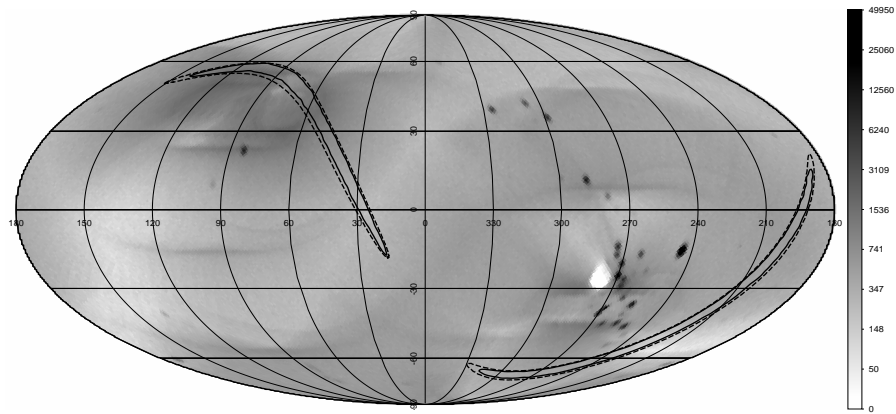}
 \end{center}
\caption{Same as figure \ref{fig:gsc1scan} but 
  from $t0$ to $t0$+10 days.
      }
\label{fig:gsc10days}
\end{figure*}

  We calculated the coverage of the 90 percent region of 
  both initial BAYESTAR \citep{2015GCN..18728...1L} and 
  refined LALInference \citep{2016GCN..18889...1L}
  GW skymaps.
  Table \ref{tab:cover} gives the GSC-covered fraction of the GW error 
  regions, for three representative time intervals.

  Figures \ref{fig:gsc1scan}, \ref{fig:gsc1day}, and \ref{fig:gsc10days}
  show the all sky X-ray images obtained by the working 6 GSC cameras in
  the first orbit (92 min),
  in 1 day and in 10 days, respectively.
  In producing these GSC images,  we did not use the ``right'' half of the
  GSC\_0 and GSC\_3, whose observing regions were covered by normal
  cameras.
  As can be seen from these images, the GSC observed about 85\% and 99\% 
  of 
  the GW error regions in the first orbit and in 1 d, respectively. 
  At ten days after the GW event, the GSC covered the whole
  of the GW error regions.

\begin{figure*}
 \begin{center}
\includegraphics[width=12cm]{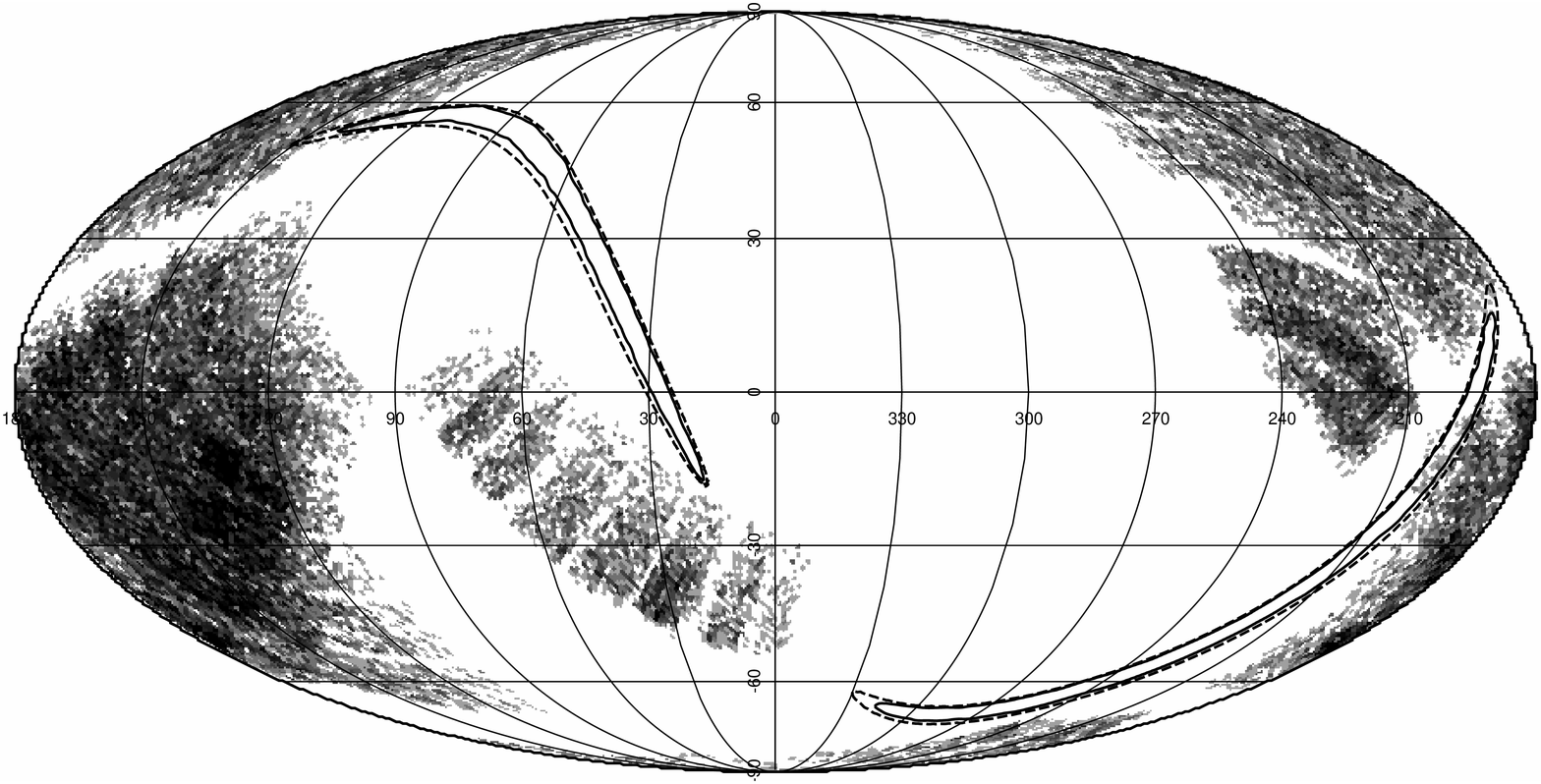}
 \end{center}
\caption{
 SSC X-ray image in 1--5 keV created only with single-pixel events 
 observed from $t0$ to $t0$+10 days.
 }
\label{fig:ssc10day}
\end{figure*}

  On the other hand, scarcely any part of the GW error regions were 
  observed by SSC.
  Figure \ref{fig:ssc10day} shows the all sky X-ray image obtained by SSC.
  Only 16.5\% of the 90 percent region of BAYESTAR map was covered.
  The observation coverage of SSC is also shown in table \ref{tab:cover}.
  Moreover, the effective exposure of SSC observation was low.
  It was 168 cm$^2$ s at most and lower than 50 cm$^2$ s in 
  the 55\% of the covered 90 percent region for 10 day observation.
  3$\sigma$ upper limit of SSC in 1-5 keV is about 0.25 
  photons cm$^{-2}$ s$^{-1}$ for the effective exposure of 168 cm$^2$ s.

\begin{table}
  \tbl{observation coverage of the GW error region.}{%
  \begin{tabular}{lrrrr}
    \hline
     map   & \multicolumn{2}{c}{BAYESTAR} & 
               \multicolumn{2}{c}{LALInference} \\
      interval 
       & \multicolumn{4}{c}{coverage\footnotemark[$*$]} \\
            & GSC & SSC & GSC & SSC \\
    \hline
     1 orbit &  84.8\% & ---    &  84.0\% & --- \\
     1d      &  99.3\% &  0.2\% &  99.2\% &  0.0\% \\
     10d     & 100.0\% & 16.5\% & 100.0\% & 14.4\% \\
    \hline
  \end{tabular}}\label{tab:cover}
  \begin{tabnote}
    \footnotemark[$*$] 
      observation coverage of 90 percent region by GSC and SSC \\
  \end{tabnote}
\end{table}

 \subsection{Event search by the nova-alert system}
 
  If a GW source emit a significant amount of X-rays,
  the MAXI nova-alert system \citep{2016PASJ...68S...1N} may find
  the emission as a transient X-ray source.
  However, the MAXI nova-alert system was not triggered 
  in any GW error regions for 1 day since the GW trigger time, 
  meaning that there was no significant variability with a typically 
  more than $3\sigma$ level.
  The nova-alert system consists of two nova-search systems to find 
  time variability of the sky, and a following alert system to 
  promptly evaluate 
  statistical significance of the variability.

\begin{figure*}
 \begin{center}
 \includegraphics[width=12cm]{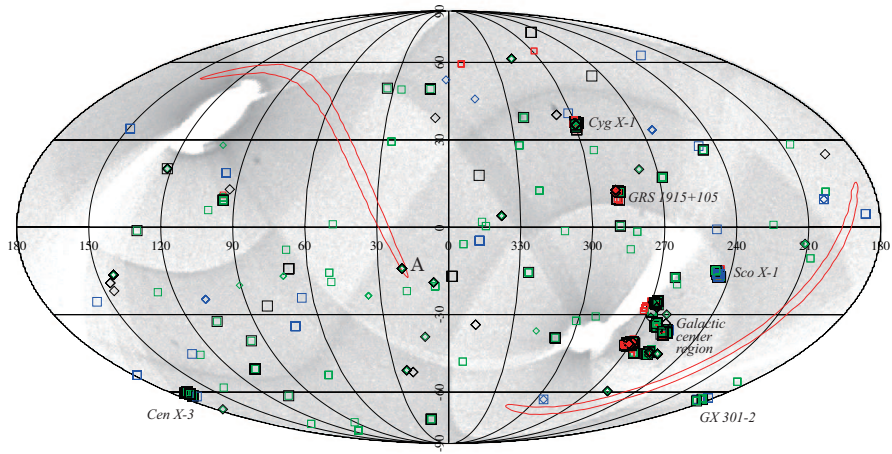}
 \end{center}
 \caption{Event positions having triggered a nova-search system of
 short-term events and long-term events.
  The black, red, green, and blue colors of the marks represent
  the triggered energy bands.
  A position with a label 'A' is the only position in the GW error region
 (see text for details). 
 The image is the 2--20 keV 1 d GSC image, and the LALInference contour is 
 shown and the BAYESTAR one is omitted.
 (Color online)
 }
\label{fig:nsdetection}
\end{figure*}

  In figure \ref{fig:nsdetection}, we plot locations of 
  events which triggered one of the two nova-search systems 
  (one with a relatively higher event threshold).
  The diamonds represent short-term events that triggered 
  from $t0-$100 s to $t0+$92 min
  in any of integrated-time bins 
  (1 s, 3 s, 10 s, 30 s, and 1-orbit ($\simeq 92$ min)).
  The squares show long-term events that
  triggered from $t0$ + 1 orbit to $t0$ + 4 orbits (in 4 orbits bin), 
  and from $t0$ + 4 orbits to $t0$ + 1 d (1 d bin).
  The black, red, green, and blue colors of the marks 
  ([see electric version]) represent the
  triggered energy bands; which are at energies of 3--10 keV, 2--4 keV, 
  4--10 keV, and 10--20 keV, respectively. 
  Chance probabilities to be triggered by background fluctuations, 
  i.e., the trigger criteria, are 
  $\le 10^{-3}$ to $10^{-4}$. 
  Except for triggers by bright known source activities as shown in 
  figure \ref{fig:nsdetection}, the system detected 106 short-term 
  triggers in 29 of 49152 sky pixel regions, 
  divided by the HEALPix library \citep{2005ApJ...622..759G}, 
  and 230 long-term triggers in 105 regions.
  Most these events are thought to be statistical noise because of 
  the relatively low criteria.

  A series of 15 events at the position denoted by '{\bf A}' 
  in figure \ref{fig:nsdetection}, however, just fall in the GW error 
  region.
  At this position, 
  the 3-10 keV and 4-10 keV 
  count rates exceeded the trigger thresholds of the nova-search system,
  from $t0+5257$ s to $t0+5260$ s 
  (in the first scan transit at the region after the GW trigger).
  This is true for time bins of
  10 s, 30 s and 1-orbit.
  But, none of the {\it detection} criteria of the alert system 
  had not been satisfied (for the {\it detection} criteria, 
  see \citet{2016PASJ...68S...1N} in more detail).

\begin{figure}
 \begin{center}
 \includegraphics[width=8cm]{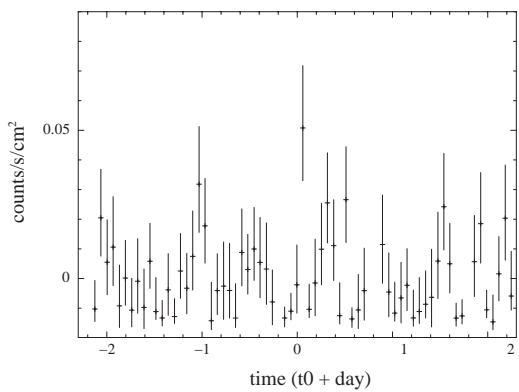}
 \end{center}
\caption{A 4--10 keV GSC light curve at the position A shown in figure 
\ref{fig:nsdetection}.

The data were obtained in each scan transit using the image fitting method 
of \citet{2016PASJ...68S..11M}.
      }
\label{fig:lc_a}
\end{figure}

  As shown later, the count-rate excess at '{\bf A}' is only 2.85$\sigma$.
  However, the source image shows a point-source-like structure.
  By assuming a constant source intensity, 
  we obtained the source position at 
  (R.A., Dec) = (19.913 deg, $-$14.480 deg) = 
  (\timeform{01h19m39s}, \timeform{-14D28'48''}) (J2000),
  and the statistical 90\% C.L. elliptical error region
  has the long and short radii of \timeform{0D.49} and \timeform{0D.46}, 
  respectively.
  The roll angle of the long axis from the north direction is 
  \timeform{174D.0} counterclockwise. 
  There is an additional systematic uncertainty of 
  \timeform{0D.1} for the above position (90\% containment radius).
  Figure \ref{fig:lc_a} shows a 4-10 keV light curve in the 1-scan 
  bin at this position.
  The source flux averaged over a scan is 
  $5.1^{+2.1}_{-1.8} \times 10^{-2}$ counts cm$^{-2}$ s$^{-1}$ 
  at the maximum,%
  \footnote{The unit of the source flux is counts cm$^{-2}$ s$^{-1}$
  instead of photons cm$^{-2}$ s$^{-1}$, because the image fitting 
  procedure simply provides the count flux and its errors, and we do not
  apply any corrections for them.
  The discrepancy between count and photon fluxes comes from the detection
  efficiency of the GSC cameras, which is more than 80\% in 4--10 keV
  energy range.
  }
  corresponding to approximately $43\pm16$ mCrab for a Crab-like spectrum 
  source.

 \subsection{upper limits of the flux}
\begin{figure}
 \begin{center}
 \includegraphics[width=8cm]{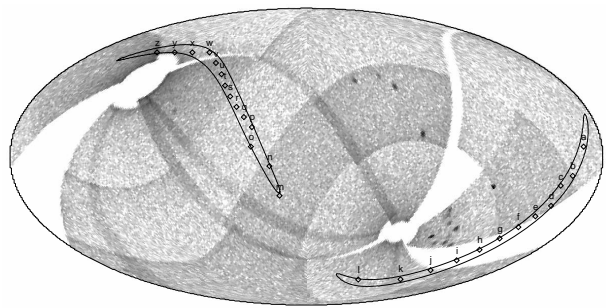}
 \end{center}
\caption{The points used to calculate the GSC signal upper limits.
  The points are selected uniformly within the GW 90\% probability 
  contours. The gray-scale is the same as
  figure \ref{fig:gsc1scan}.
      }
\label{fig:pos}
\end{figure}

  We evaluate the upper limits of the X-ray flux associated
  with GW151226 by the following procedure.
  As shown in figure \ref{fig:pos}, we first selected 26 points 
  representing the observed region.
  Then we
  counted the photons in the circular regions with the radii of 
  \timeform{2D.0}, which is large enough to cover the point spread 
  function.
  The 1$\sigma$ fluctuation of the background is taken as $\sqrt{n}$,
  where $n$ is the observed counts within the circular region.
  Next, we calculated the effective exposure $a$, which has 
  the dimension of area $\times$ time.
  Then we regarded $f \equiv  \, \sqrt{n} / a $ as 1$\sigma$ upper limit
  of the flux at the point.
  Since the observation condition 
  (camera, position in the detectors, and exposure)
  depends on the position,
  the upper limits are different.
  The 3$\sigma$ upper limits of each point for the observations 
  of one orbit, one day, and ten days are shown in table \ref{tab:ul}.
  This upper limit for one orbit is comparable to the GSC upper limit for
  GW150914 (0.12 photons cm$^{-2}$ s$^{-1}$, \cite{gw150914.maxi}),
  and also with the 4-10 keV flux of the event {\bf A} discussed in section 3.2.

\begin{table*}
  \tbl{observation coverage of the GW maps.}{%
  \begin{tabular}{c c crrl c crrl c crrl}
    \hline
& & \multicolumn{4}{c}{one orbit} & &
  \multicolumn{4}{c}{one day} & &
  \multicolumn{4}{c}{ten days}   \\
  \cline{3-6} \cline{8-11} \cline{13-16}
pt\footnotemark[$*$] & RA, Dec\footnotemark[$\dagger$]  
   & cam\footnotemark[$\ddagger$] 
      & cnt\footnotemark[$\S$]
       & exp\footnotemark[$\|$] & U.L.\footnotemark[$\#$] &
   & cam\footnotemark[$\ddagger$] 
      & cnt\footnotemark[$\S$]
       & exp\footnotemark[$\|$] & U.L.\footnotemark[$\#$] &
   & cam\footnotemark[$\ddagger$] 
      & cnt\footnotemark[$\S$]
       & exp\footnotemark[$\|$] & U.L.\footnotemark[$\#$] 
    \\
    \hline
a & 191.25, 4.93 &  4     &  26   &   87 & 0.17   &
& 4     &  373  &  1329 & 0.04  & & 4,5    &   5901  &  19222 & 0.012 \\
b & 196.87, -9.44 & ---    & ---   &   0  &  ---   &
& 4,5   &  481  &  1012 & 0.06  & & 4,5    &   7414  &  23602 & 0.011 \\
c & 202.50, -14.32 &  4,5   &  51   &  155 & 0.14   &
& 4,5   &  775  &  2209 & 0.04  & & 4,5    &   7187  &  23289 & 0.011 \\
d & 202.50, -24.46 &  4,5   &  49   &  164 & 0.13   &
& 4,5   &  599  &  1999 & 0.04  & & 4,5    &   6291  &  21129 & 0.011 \\
e & 208.12, -29.83 &  4,5   &  39   &  145 & 0.13   &
& 4,5   &  681  &  2073 & 0.04  & & 4,5    &   6304  &  20276 & 0.012 \\
f & 213.75, -35.50 & ---    & ---   &   0  &  ---   &
& 4,5   &  663  &  2004 & 0.04  & & 4,5    &   5801  &  18814 & 0.012 \\
g & 219.37, -41.61 & ---    & ---   &   0  &  ---   &
& 4,5   &  532  &  1542 & 0.04  & & 4,5    &   4913  &  15456 & 0.014 \\
h & 225.00, -47.95 & ---    & ---   &   0  &  ---   &
& ---   &  ---  &     0 &  ---  & & 5      &   2761  &   8620 & 0.018 \\
i & 232.46, -54.15 & ---    & ---   &   0  &  ---   &
& 5     &  268  &   404 & 0.12  & & 2,5    &   2624  &   4789 & 0.030 \\
j & 242.89, -60.25 &  2     &  27   &  96  & 0.16   &
& 2     &  141  &   332 & 0.11  & & 2      &   5620  &  10958 & 0.021 \\
k & 258.49, -66.26 &  2     &  39   & 126  & 0.15   &
& 2     &  500  &  1404 & 0.05  & & 2      &   6371  &  18821 & 0.013 \\
l & 303.75, -66.44 &  2     &  29   & 142  & 0.11   &
& 2     &  656  &  2124 & 0.04  & & 2      &   5619  &  18788 & 0.012 \\
m & 16.87, -19.31 &  4,5   &  47   & 171  & 0.12  &
& 4,5   &  672  &  2201 & 0.04  & & 4,5,7  &   4734  &  14245 & 0.014 \\
n & 22.50, -4.63 &  4,5   &  36   & 152  & 0.12   &
& 4,5   &  737  &  1968 & 0.04  & & 4,5,7  &   3449  &  11272 & 0.016 \\
o & 33.75, 4.93 &  4     &  19   &  91  & 0.14   &
& 4     &  336  &  1347 & 0.04  & & 4,7    &   3755  &   9504 & 0.019 \\
p & 33.75, 14.63 &  4     &  25   &  88  & 0.17   &
& 4     &  368  &  1308 & 0.04  & & (3),4  &   2819  &   7463 & 0.021 \\
q & 39.37, 19.63 &  4     &  31   &  83  & 0.20   &
& 4     &  402  &  1228 & 0.05  & & (3),4  &   2244  &   4967 & 0.029 \\
r & 45.00, 24.79 &  4     &  27   &  76  & 0.20   &
& 4     &  488  &  1125 & 0.06  & & (3),4  &   1368  &   2651 & 0.040 \\
s & 50.62, 30.17 &  4     &  39   &  68  & 0.27   &
& 4     &  369  &   603 & 0.10  & & (3),4  &   369   &    603 & 0.10 \\
t & 56.25, 35.87 &  3     &  70   &  64  & 0.39   &
& 3     &  1067 &   930 & 0.11  & & 3      &   22135 &  12678 & 0.040 \\
u & 61.94, 42.01 &  3     &  89   &  70  & 0.40   &
& 3     &  1065 &   861 & 0.11  & & 0,(3)  &   54    &    229 & 0.10 \\
v & 70.83, 48.34 &  3     &  136  &  77  & 0.45   &
& 3     &  1167 &   784 & 0.13  & & 0,(3)  &   6593  &   8967 & 0.027 \\
w & 82.70, 54.53 &  0     &  67   & 151  & 0.16   &
& 0     &  737  &  1710 & 0.05  & & 0      &   16081 &  23736 & 0.016 \\
x & 97.30, 54.53 &  0     &  69   & 172  & 0.14   &
& 0     &  1358 &  2456 & 0.05  & & 0,(3)  &   15237 &  19872 & 0.019 \\
y & 112.38, 54.53 &  0     &  126  & 192  & 0.18   &
& 0     &  2264 &  3060 & 0.05  & & 0,(3)  &   9097  &  12227 & 0.023 \\
z & 127.46, 54.53 &  0,(3) &  122  & 179  & 0.18   &
& 0,(3) &  1196 &  1902 & 0.05  & & 0,(3)  &   2145  &   4066 & 0.030 \\
    \hline
  \end{tabular}}\label{tab:ul}
  \begin{tabnote}
    \footnotemark[$*$] 
      point ID shown in figure \ref{fig:pos} \\
    \footnotemark[$\dagger$]
      position of the point in J2000 coordinates \\
    \footnotemark[$\ddagger$]
      ID of the cameras which observed the point.
      Camera number in parenthesis was not used \\
    \footnotemark[$\S$]
      observed counts in a circular region of 2 degree radius \\
    \footnotemark[$\|$]
      effective exposure which is the cumulative effective area
      for the exposure in the unit of cm$^2$ s \\
    \footnotemark[$\#$]
      3$\sigma$ upper limit in 2--20 keV in the unit of photons cm$^{-2}$ s$^{-1}$
  \end{tabnote}
\end{table*}
\label{ss:obs}

\section{Discussion}

X-ray counterparts of GW events would be detected as 
short gamma-ray bursts
(SGRBs; 
\cite{2007PhR...442..166N,2014ApJ...796...13N}).
Therefore we compare the flux upper limits in the previous section to
expected X-ray fluxes of SGRBs.

First, we compare the upper limits of one scan with a typical
flux of an extended emission \citep{2006ApJ...643..266N}, 
which is long ($\sim$ 100 s) X-ray emission
after the short hard pulse of a SGRB.
The 3$\sigma$ GSC upper limits of a scan for GW151226
range from 0.11 to 0.45 photons cm$^{-2}$ s$^{-1}$ in 2--20 keV. 
If we assume a power-law spectrum with a photon index of 2.0,
which is typical of extended emissions of SGRBs
\citep{2015MNRAS.452..824K},
the corresponding energy fluxes are 0.9--3.7 
$\times 10^{-9}$ ergs cm$^{-2}$ s$^{-1}$ in 2--20 keV.
The observed peak fluxes of extended emissions range from
$10^{-8}$ to $5\times10^{-6}$ ergs cm$^{-2}$ s$^{-1}$ in 15--350 keV
\citep{2015MNRAS.452..824K}, corresponding to the fluxes of
$2\times10^{-9}$ to $10^{-5}$ ergs cm$^{-2}$ s$^{-1}$ in 2--20 keV,
when the photon index is 2.0.
Therefore MAXI/GSC is sensitive to extended emissions with a typical flux.

Next, we consider the case of a short pulse.
If the X-ray emission lasts only for 1 sec, the result changes as follows.
We need $\sim$ 10 photon counts for a 3$\sigma$ detection of a source.
If a GSC camera detects 10 photons in a second and the effective area
toward the source is 1 cm$^{2}$ in 2--20 keV%
\footnote{The effective area toward a source continuously changes
during a scan transit. The duration where the effective area $>$ 1 cm$^{2}$
contains about 70\% and 85\% of the duration of a scan transit
for the observation of single camera and two cameras, respectively.}
at that time 
(equivalent to the photon flux of 10 photons cm$^{-2}$ s$^{-1}$),
it corresponds to the
energy flux of 1 $\times 10^{-7}$ ergs cm$^{-2}$ s$^{-1}$.
Here we assume a power-law spectrum with a photon index of 1.0, which is
typical of short pulses of SGRBs
\citep{2015MNRAS.452..824K}.
In the same way,
in order to detect a burst with a duration of 0.1 sec, the 
2--20 keV flux should be more than
1 $\times$ 10$^{-6}$ ergs cm$^{-2}$ s$^{-1}$.
The observed peak fluxes of initial spikes range from $5\times10^{-8}$
to $10^{-5}$ ergs cm$^{-2}$ s$^{-1}$
\citep{2013MNRAS.428.1623B,2015MNRAS.452..824K}, 
corresponding to the flux of
$3\times10^{-8}$ to $5\times10^{-6}$ ergs cm$^{-2}$ s$^{-1}$ in 2--20 keV,
and thus not all of them can be detected by MAXI/GSC.

The distance to the source of GW151226 was estimated as 
440 $^{+180}_{-190}$ Mpc
\citep{2016PhRvL.116x1103A}.
On the other hand \citet{2014MNRAS.442.2342D} reported the 
average redshift of SGRB is 0.85,
which is $\sim$ ten times as far as GW151226.
The lowest redshift in the samples of \citet{2014MNRAS.442.2342D} and
\citet{2015MNRAS.452..824K} are 0.122 and 0.125, respectively,
and they are comparable to GW151226.
The fact suggests that MAXI/GSC will be able to detect emissions from 
GW events, especially extended emissions,
if the GW events are at the distance of GW151226 or nearer and
they are accompanied by SGRBs.

Finally we note that MAXI is capable of detecting bright
hard GRBs even out of the field of view (FOV).
These GRBs penetrated the shields and the collimators of GSC and
deposit the energy in the detectors.
Although the processes of the energy deposit, the effective area, and 
the detection efficiency of these events are different from those 
of in-FOV events,
the detection limits of out-of-FOV events are similar by chance
to those of in-FOV events.
For example, MAXI detected three out-of-FOV events, GRB 120816B, 
GRB 140219A \citep{2014GCN..15882...1S}, and GRB 160625B,
whose peak fluxes range from $10^{-4}$ to $10^{-3}$ erg cm$^{-2}$ s$^{-1}$
in 20 keV--10 MeV
\citep{2012GCN..13676...1G,2014GCN..15870...1G,2016GCN..19604...1S}
and corresponding energy fluxes in 2--20 keV range from $10^{-7}$ to 
$10^{-6}$.%
\footnote{The $E_{\rm peak}$ of SGRB is about 1 MeV. It makes the flux
difference between two energy bands by a factor of 10$^{-3}$}
The observed duration of short hard GRB 120816B by MAXI was about 0.8 s,
which was consistent with the result of \citet{2012GCN..13676...1G}.
If short hard GRBs are really associated with some of GW events,
MAXI/GSC may detect emissions of them.
Although MAXI cannot determine the position of such out-of-FOV GRBs,
the position may be determined with the triangulation method using
the arrival time of the GRB.

\begin{ack}
This research has made use of the MAXI data provided by RIKEN, JAXA 
and the MAXI team.
This research was supported by JSPS KAKENHI
Grant Number 24740186, 24540239, 16K05301, 17K05402.
\end{ack}


\bibliographystyle{aa}
\bibliography{ref}

\end{document}